\documentclass[prd,aps]{revtex4}
\usepackage[utf8]{inputenc}
\usepackage{amssymb}
\usepackage{amsmath}
\usepackage{cancel}
\usepackage{mathtools}
\usepackage{graphicx}
\usepackage{epsfig}
\usepackage{bm}
\usepackage{amsfonts}
\usepackage{color}

\usepackage{epsf}
\usepackage{epsfig,graphicx}
\begin{document}
\title{Cosmological evolution in a two-brane warped geometry model }
\author{Sumit Kumar, Anjan A Sen}
\affiliation{Center For Theoretical Physics\\Jamia Millia Islamia, New Delhi 110025, India}
\email{sumit@ctp-jamia.res.in, aasen@jmi.ac.in}
\author{Soumitra SenGupta}
\affiliation{Department of Theoretical Physics\\Indian Association for the Cultivation of Science\\Kolkata 700032, India}
\email{tpssg@iacs.res.in}

\begin{abstract}
We study  an  effective 4-dimensional scalar-tensor field theory, originated from an underlying brane-bulk warped geometry, to explore the scenario of inflation.
It is shown that the inflaton potential naturally emerges from the radion energy-momentum tensor which in turn results into  an inflationary model of the Universe on the visible brane 
that is consistent with the recent results from the Planck's experiment. The dynamics of modulus stabilization from the inflaton rolling condition is demonstrated. The implications
of our results in the context of recent BICEP2 results are also discussed.  

\end{abstract}
\maketitle
\section{Introduction}

The standard cosmological paradigm, while successful in describing our observable Universe, is plagued with horizon and flatness problems. 
Moreover, despite being able to explain the large scale structure formation due to some seed fluctuations in our Universe, standard cosmology fails to provide a mechanism that can produce 
such seed fluctuations. Inflationary models are at present the only way to provide solutions for these shortcomings in standard cosmology \cite{inf}. According to this paradigm ,
the Universe at an early epoch experienced an exponentially rapid expansion for a very brief period 
due to some apparently repulsive gravity-like force. Such a scenario not only can successfully address  the horizon and flatness problems but at the same time, provides a theoretical set 
up to produce the primordial fluctuations which later may act as a seed for large scale structure formation in the Universe. Amazingly the predicted primordial 
fluctuations in any inflationary model \cite{pert} can be tested accurately through the measurement of  temperature anisotropies in the Cosmic Microwave Background Radiation as recently 
done by Planck experiment \cite{planck}. The construction of a  viable models for inflation, which are consistent with cosmological observations like Planck experiment, therefore is of
utmost importance and is a subject of study of the present work.

Among various models for inflation, the  models with extra dimensions have been discussed by many authors \cite{extrainfla}. Such models are independently considered in particle 
phenomenology due to their promise of resolving the well-known naturalness/fine tuning  problem in connection with stabilising the mass of Higgs boson against 
large radiative corrections \cite{solnhier}.

In this context, the 5-dimensional warped geometry model due to Randall and Sundrum ( RS ) \cite{RS} is very successfull  in offering a proper resolution to the naturalness problem 
without
incorporating any intermediate scale other than Plank/quantum gravity scale. The radius  associated with the extra dimension in this model ( known as  RS modulus ) acts as 
a parameter in the effective 4-dimensional theory and from a cosmological point of view, such a  modulus can be interpreted as a scalar field which, due to it's time evolution,
may drive the scale factor of our universe before getting stabilized  to a desired value.
The well-known methodology to extract an effective or induce theory on a 3-brane from a 5-dimensional warped geometry model is demonstrated in \cite{shirkoy}
where using the  Gauss-Codazzi
equation with appropriate junction condition in a two-brane warped geometry model and implementing a  perturbative expansion 
in terms of the brane-bulk  curvature ratio, the effective Einstein's equation is obtained on a lower dimensional hypersurface. 
This eventually results in the form of a scalar-tensor  gravity theory in our 
brane ( known as visible brane ) \cite{shirkoy}.

Here  we try to explore the role of such a scalar-tensor theory 
in stabilizing the modulus of the bulk geometry as well as to generate inflation in the visible brane which is consistent with the results 
obtained by Planck. We show that an effective potential for the modulus field ( often called radion ) automatically emerges from the construction of the model.  
The stabilisation requirements put further  constrains on various parameters of the  modulus/scalar potential. 
After deriving these constraints we  study the cosmological evolution in the 
Einstein frame in presence of such a potential and show that it gives a viable model for inflation with required number of e-folds (to solve the horizon and flatness problem)  and also gives a 
primordial fluctuations which is perfectly consistent with the Planck results. The inflation is shown to end with the modulus attaining it's stable value.
Hence our set up not only provides a mechanism to stabilise the modulus in the bulk but also provides a viable 
model for inflation which is consistent with the recent observational results.

The structure of the paper is as follows: in section II, we briefly review the Shiromizu and Koyama set up \cite{shirkoy} for the low energy effective gravity in curved branes; in section III, we investigate the constraints on the potential that is necessary for moduli stabilisation; in section IV, we describe the inflationary behaviour in our model and constrain it using the recent result from the Planck experiment; in section V, we briefly comment about the recent BICEP2 results; we end with conclusions in section VI.

\section{Low energy effective gravity in presence of curved branes}

We start with a configuration which contains two 3-branes embedded in a five dimensional  $z_{2}$ symmetric ADS spacetime containing a bulk cosmological constant $(\Lambda_{5})$. 
The branes are located at two orbifold fixed points. One has  positive tension and is placed at $y=0$ in the fifth dimension (the "hidden brane") while the other has negative tension and 
is placed at $y=r\pi$ (the "visible brane"), $r$ being the distance between the two branes. 

\noindent
Next we assume the five dimensional action as \cite{RS}:

\begin{equation}
{\cal S} = \frac{1}{2\kappa^2}\int d^{5}x \sqrt{-g} \left(R + \frac{12}{l^2}\right) - \sum _{i = 1,2} {\cal{V}}_{i} \int d^{4}x \sqrt{-g^{i}} + \sum_{i=1,2}\int d^{4}x \sqrt{-g^{i}} {\cal L}^{i}_{matter} .
\end{equation}

\noindent
Here $\kappa^2$ is the five dimensional gravitational constant, $l$ is the bulk curvature radius which is related to the bulk cosmological 
constant as $l = \sqrt{ \frac{-3}{{\kappa^2}\Lambda_{5}}}$.  ${\cal V}_{1}$ and ${\cal V}_{2}$ are the tensions of the hidden  and the visible branes. The 5D line element is taken as:

\begin{equation}
ds^{2} = e^{2\phi} dy^2 + g_{\mu\nu}(y, x^{\mu}) dx^{\mu}dx^{\nu}.
\end{equation}

Fixing the bulk curvature  scale to be $L$, we define a parameter $\epsilon = (\frac{L}{l})^2$ and assume $\epsilon <<1$ which is legitimate as the scale of the cosmological evolution in 
brane is considerably smaller than the Planck's scale namely the natural scale for the bulk curvature. This ensures that the  classical solutions of the effective Einstein's equation 
can be trusted. One can perturbatively expand the extrinsic curvature of the brane at fixed $y$  in terms of $\epsilon$. At the zeroth order one retrieves  the RS model with the
corresponding brane tensions:

\begin{equation}
\frac{1}{l} = \frac{1}{6} \kappa^2 {\cal{V}}_{1} = -\frac{1}{6} \kappa^2 {\cal{V}}_{2}. 
\end{equation}
The Einstein tensor can be calculated from the given action (see \cite{shirkoy} for detail derivation) and 
in the first order, one can get the effective Einstein's equation on the visible brane as:

\begin{equation}
G^{\mu}_{\nu} = \frac{{\kappa}^2}{l}\frac{{T_2}^{\mu}_{\nu}}{\Phi}+\frac{{\kappa}^2}{l}\frac{(1+\Phi)^2}{\Phi}{T_1}^{\mu}_{\nu}+\frac{1}{\Phi}\left(D^{\mu}D_{\nu}\Phi - \delta^{\mu}_{\nu}D^2\Phi\right)+\frac{\omega(\Phi)}{{\Phi}^2}\left({D^{\mu}\Phi}D_{\nu}\Phi - \frac{1}{2}\delta^{\mu}_{\nu}({D_\alpha\Phi}D^{\alpha}\Phi)\right), \label{1}
\end{equation}

\noindent
where $\Phi = \exp^{2d_{0}(x)/l} - 1$ and $\omega(\Phi) = -\frac{3}{2}\frac{\Phi}{1+\Phi}$. Here $d_{0} (x) = \int_{0}^{r\pi} dy e^{\phi(y,x)}$ is the proper distance between the two branes
and is the modulus field in the effective 4-dimensional theory. 
$T_{1 \nu} ^{\mu}$ and $T_{2, \nu}^{\mu}$ are the respective energy momentum tensors in the hidden and the visible branes.

\noindent
If we now assume that the two branes are endowed with only cosmological constants i.e ${T_2}^{\mu}_{\nu}={\Lambda_2}\delta^{\mu}_{\nu}$ and ${T_1}^{\mu}_{\nu}={\Lambda_1}\delta^{\mu}_{\nu}$ then
for a spatially flat FRW metric with scale factor $a(t)$, the Einstein equations in the visible brane are  given by:

\begin{eqnarray}
3H^2 &=& \frac{1}{2}\omega(\Phi)\left(\frac{\dot{\Phi}}{\Phi}\right)^2-3H\left(\frac{\dot{\Phi}}{\Phi}\right)-\frac{\kappa^2}{l}\frac{\Lambda_2}{\Phi}(1+(1+\Phi)^2\left(\frac{\Lambda_1}{\Lambda_2}\right))\label{e3}\\
2\dot{H}+3H^2 &=&-\frac{1}{2}\omega(\Phi)\left(\frac{\dot{\Phi}}{\Phi}\right)^2-\frac{\ddot{\Phi}}{\Phi}-2H\left(\frac{\dot{\Phi}}{\Phi}\right)-\frac{\kappa^2}{l}\frac{\Lambda_2}{\Phi}(1+(1+\Phi)^2\left(\frac{\Lambda_1}{\Lambda_2}\right))\label{e4},\nonumber\\
\end{eqnarray}
where $H= \frac{\dot{a}}{a}$ is the Hubble parameter. The term
\begin{equation}
U(\Phi)=-\frac{\kappa^2}{l}\Lambda_2\left(1+(1+\Phi)^2\left(\frac{\Lambda_1}{\Lambda_2}\right)\right) \label{4}
\end{equation}
can be interpreted as the potential for the scalar field $\Phi$ in this model. 

\section{Constraints on the Form of the potential}

To study the dynamical evolution of our system, it is convenient to write the equations in the Einstein frame which can be obtained using the following conformal transformations:

\begin{eqnarray}
\tilde{g}_{\mu\nu} &=& \Phi g_{\mu\nu} \nonumber \\
\tilde{a}^2 &=& \Phi a^2  \\
d\tilde{\tau}^2 &=& \Phi d{\tau}^2 \nonumber
\end{eqnarray}  
We use a field redefinition $\Phi \longrightarrow \psi$, such that 
\begin{equation}
\left(\frac{d\psi}{d\Phi}\right)^2=\frac{3}{4}\frac{1}{\Phi^2(1+\Phi)}.
\end{equation}

\noindent
On solving the above equation we arrive at,
\begin{equation}
\psi = \pm\frac{\sqrt{3}}{2}\ln{\left|\frac{\sqrt{1+\Phi}-1}{\sqrt{1+\Phi}+1}\right|}+c_{1}\nonumber
\end{equation}
where $c_{1}$ is a constant of integration.
We can also write Jordan frame field $\Phi$ in terms of field $\psi$ in the Einstein frame as
\begin{equation}
\Phi=\frac{4\alpha}{(1-\alpha)^2}
\end{equation}
where
\begin{eqnarray}
\alpha &=& Ke^{\pm\frac{2}{\sqrt{3}}\psi} \label{e15}
\end{eqnarray}
with  $K=e^{\mp\frac{2}{\sqrt{3}}{c_{1}}}$ 

Potential $V(\psi)$ in Einstein frame is now related to Potential $U(\Phi)$ in Jordan frame as\cite{polarski},
\begin{equation}
V(\psi)=\frac{U(\Phi)}{2F^2(\Phi)}.
\end{equation}
We further define two parameters
\begin{eqnarray}
A &=& \frac{\kappa^2\Lambda_2}{2l}, \nonumber\\
B &=& \frac{\Lambda_1}{\Lambda_2}.
\end{eqnarray}

In terms of the parameters $A,B$ and $K$, $V(\psi)$ has the form:
\begin{equation}
V(\psi)=-A\left(\frac{(1-Ke^{\pm\frac{2}{\sqrt{3}}\psi})^4+(1+Ke^{\pm\frac{2}{\sqrt{3}}\psi})^4B}{16K^2e^{\pm\frac{4}{\sqrt{3}}\psi}}\right) \label{e11}
\end{equation}
Moreover with these transformations, the Einstein equations \eqref{e3}, \eqref{e4} simplifies to
\begin{eqnarray}
3\mathcal{H}^2 = \dot{\psi}^2 + 2V(\psi) \\ 
2\mathcal{H} + 3\mathcal{H}^2 = -\dot{\psi}^2 + 2V(\psi).
\end{eqnarray}

There are several restrictions on the form of the potential so that the model can simultaneously address the following issue :\\  
Firstly to achieve the stabilisation of the brane motion, the potential should have a minimum and field value at this minimum should be 
non-zero to avoid any brane collision. 
Secondly, the  potential should also satisfy the necessary slow-roll conditions  to trigger the inflation on the visible brane and 
finally the spectrum of the primordial fluctuations produced in this case should also be consistent with its recent measurement by Planck experiment. We address these issues one by one
to ascertain the viability of the model.

First let us examine the extremum of the potential  given by  equation (11). The equation $\frac{dV}{d\psi} = 0$ gives the following set of  conditions:

\begin{eqnarray}
e^{-2\beta\psi_{min}} = 0\\
Ke^{\beta\psi_{min}} = -1\\
Ke^{\beta\psi_{min}} = 1\\
\psi_{\pm}=\frac{1}{\beta}\ln{\left[\frac{\left(\frac{1-B}{1+B}\right)\pm\sqrt{\frac{(1-B)^2}{(1+B)^2}-1}}{K}\right]} 
\end{eqnarray}

\noindent 
where $\beta = \pm \frac{2}{\sqrt{3}}$. The first and the third conditions result the corresponding $\Phi$ in the Jordan frame to be  either infinity or zero. 
Neither of these are acceptable as they imply infinite or zero separation between the two branes. 
The second condition is not possible as $K$ is strictly positive. 
So the acceptable $\psi_{min}$ is given by equation (20). Also it is easy to check that $-1<B<0$ is necessary in order to have a real $\psi_{min}$. Now if one 
further calculates $\frac{d^2 V}{d\psi^2}$, one gets

\begin{equation}
\frac{d^2 V}{d\psi^2} (\psi = \psi_{min}) = 2\beta^2 \frac{AB}{1+B}.
\end{equation}

\noindent
In order to have a minimum of the potential at $\psi=\psi_{min}$, one further needs $A < 0$.
This, together with the condition on $B$,  implies that $\Lambda_{2}$ should be negative and $\Lambda_{1}$ should be positive which is similar to the 
Randall-Sundrum set up of warped geometry. 

\noindent
Further, it is easy to show that at the minimum, 

\begin{equation}
V_{min} = -\frac{AB}{(1+B)} < 0.
\end{equation}

\noindent
But  the cosmological observations actually is consistent with a de-Sitter Universe. Hence we need to add an uplifting term  $V_{1}$ in our potential with the 
condition such that

\begin{equation}
V_{1} > \frac{AB} {(1+B)}.
\end{equation}

This feature is similar to the de-Sitter lifting by fluxes in KKLT model  \cite{KKLT} where the supersymmetry preserving ADS minima is lifted to a de-Sitter one from the energy of the 
background fluxes of higher form tensor fields in type IIB string-based $N =1$ supergravity model in presence of brane and anti-brane. Presence of anti-brane breaks the supersymmetry
giving rise to a positive definite vacuum energy by compensating the ADS value of the scalar potential at the minimum. The mechanism here is to generate some extra energy from 
background fluxes which show up in the scalar potential.  

\noindent We should stress that adding this uplifting term does not disturb the process of radion stabilisation and the subsequent solution of the hierarchy problem.

So the final form of the potential is 

\begin{equation}
V(\psi)=-A\left(\frac{(1-Ke^{\beta\psi})^4+(1+Ke^{\beta\psi})^4B}{16K^2e^{2\beta\psi}}\right) + V_{1},
\end{equation}

\noindent
where $A < 0$, $-1<B<0$. 
\begin{figure}[ht]
{\centerline{\epsfig{file=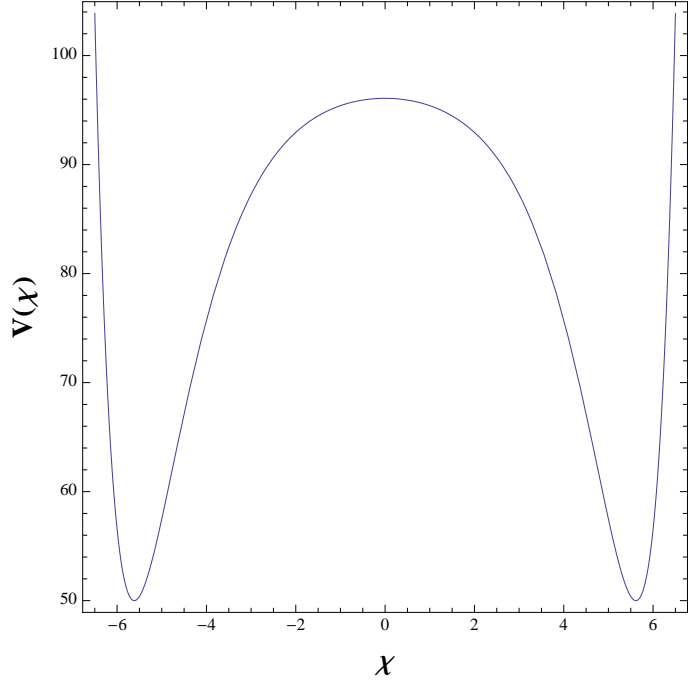,width=9cm,height=9cm,angle=0}}}
\caption{Behaviour of the potential for $B=-0.96$, $K =1$.  $A$ and $V_{1}$ are chosen to be $-1$ and $49$ in units of $\frac{\kappa^{2} \Lambda_{2}}{2l}$. Here $\chi$ is shown in units of $M_{pl}$ and $V(\chi)$ in units of $M_{pl}^4$.}
\label{fig:a_of_t}
\end{figure}

\section{Inflation}

We now study the inflationary solution induced by the  potential as given in equation (24). We define two new variables $\chi = \sqrt{2}M_{pl}\psi$ and $V(\chi)=2M_{pl}^2V(\psi)$, where $M_{pl}$ is the reduced Planck mass. The form of the potential $V(\chi)$ is shown in figure 1. As $\chi$ rolls over from the flat part near the region $\chi =0$ towards the minimum at the either 
side, inflation continues to occur.

The equations (15) and (16) now become,
\begin{eqnarray}
3\mathcal{H}^2 = \frac{1}{M_{pl}^2}\left(\frac{\dot{\chi}^2}{2}+V(\chi)\right) \label{e34}\\
2\dot{\mathcal{H}}+3\mathcal{H}^2 = \frac{1}{M_{pl}^2}\left(-\frac{\dot{\chi}^2}{2}+V(\chi)\right) \label{e35}
\end{eqnarray}

We further define slow roll parameters as\cite{pert},
\begin{equation}
\epsilon_V = \frac{M_{pl}^2}{2}\left(\frac{1}{V(\chi)}\frac{dV(\chi)}{d\chi}\right)^2 
\end{equation}
and
\begin{equation}
\eta_V = M_{pl}^2\left(\frac{1}{V(\chi)}\frac{d^2V(\chi)}{d\chi^2}\right). \label{e37}
\end{equation}

\noindent
Inflation takes place when $\epsilon << 1$ and $|\eta|<<1$ which are also called the slow-roll  conditions. 
The inflation ends when any one of these conditions breaks down. In figure (2), we show the behaviour of the slow-roll parameters for the potential shown in figure (1). The inflation 
ends at $\chi_{end} \sim 5.52 M_{pl}$ and hence it ends before the scalar field settles down at its minima at $\chi_{min} \sim 5.61 M_{pl}$.

\begin{figure}[ht]
{\centerline{\epsfig{file=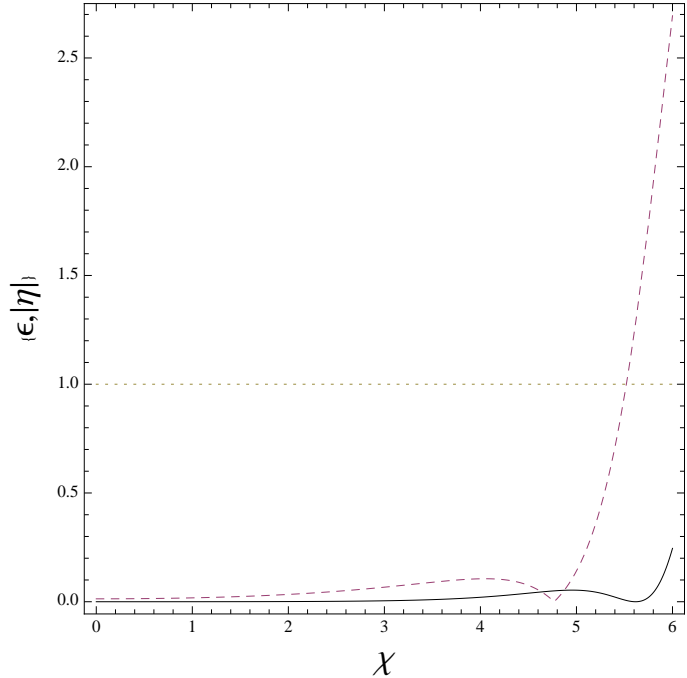,width=9cm,height=9cm,angle=0}}}
\caption{Behaviour of the slow-roll parameters. The parameters are chosen to be same as in figure (1). The dashed line is for $\eta$ and the solid line is for $\epsilon$.}
\label{fig:a_of_t}
\end{figure}

The total amount of inflation is measured through the number of efolding $N$, defined as \cite{pert}

\begin{equation}
N = ln \frac{a(t_{e})}{a(t)} =  \int ^{t_{f}} _{t} H dt
\end{equation}

To solve the flatness problem in  standard cosmology, we need at least 70 efolds of inflation. For the potential shown in figure (1) and with $\chi_{end} \sim 5.52 M_{pl}$, the required initial value of $\chi$ 
is $\chi_{ini} \sim 1.02 M_{pl}$ to achieve the desired  number of efolding. The figure (1) clearly depicts that at such a $\chi_{ini}$, the field is initially displaced slightly from the flat part of 
the  potential. After that it slowly rolls down and  one gets enough number of efolds  before it finally settles at the minimum of the potential $V(\chi)$.

The relevant observational quantities related to the spectrum of the primordial fluctuations are \cite{pert,planck}

\begin{eqnarray}
r &\approx& 16\epsilon_{V} \nonumber \\
n_s &\approx& 1 - 6\epsilon_V + 2\eta_V \\
A_s &\approx& \frac{V(\chi)}{24\pi^2M_{pl}^4\epsilon_V},\nonumber
\end{eqnarray}

\noindent
where $r$ is the tensor to scalar ratio, $n_s$ is the scalar spectral index and $A_s$ is amplitude of the scalar fluctuation. 
To comply with our purpose all these quantities here must  be calculated at the time of Hubble exit $k_{*} = a_{*}H_{*}$. When the scale $k_{*}$ leaves the 
Hubble radius, the number of efolding before the end of inflation, $N_{*}$, is given by

\begin{equation}
N_{*}  \approx \frac{1}{M_{pl}^2} \int ^{\chi_{e}}_{\chi_{*}} d\chi \frac{V}{V_{\chi}}.
\end{equation}

So the quantities $r$, $n_{s}$ and $A_{s}$ should be estimated at $N_{*}$. The value of $N_{*}$ depends crucially on the reheating mechanism. For reasonable 
inflationary models, one can show that $50< N_{*}<60$.  In our case we take $N_{*} = 55$ which is consistent with the Planck's analysis.

The constraints obtained by the Planck's measurements are as follows \cite{planck}: 

\begin{eqnarray}
r &<& 0.11 \nonumber \\
n_s &=& 0.9603\pm 0.0073\\
ln(10^{10}A_{s}) &=& 3.089^{+0.024}_{-0.027}. \nonumber
\end{eqnarray}

\begin{center}
\begin{figure*}[t]
\begin{tabular}{c@{\qquad}c}
\epsfig{file=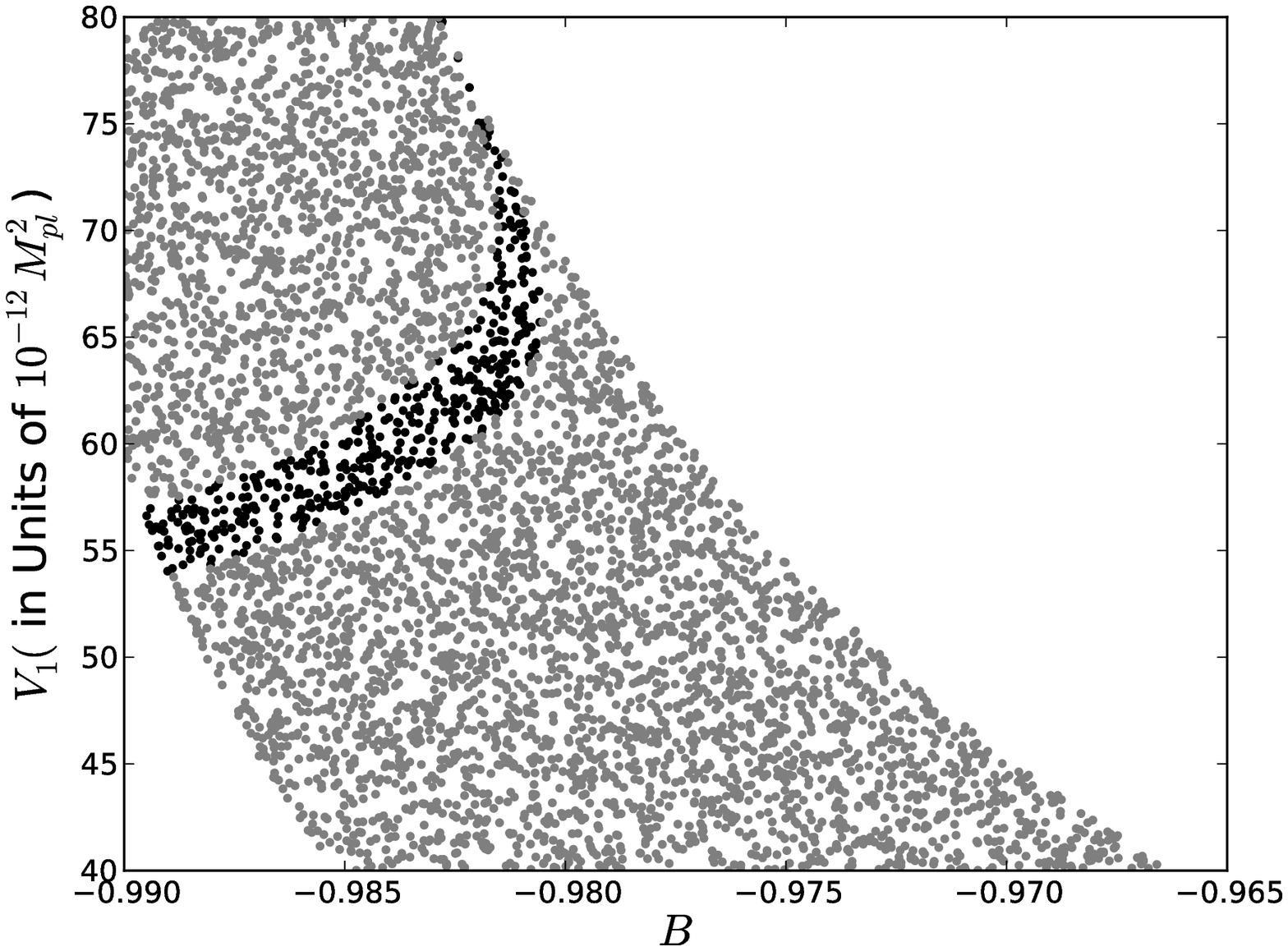,width=7 cm}&
\epsfig{file=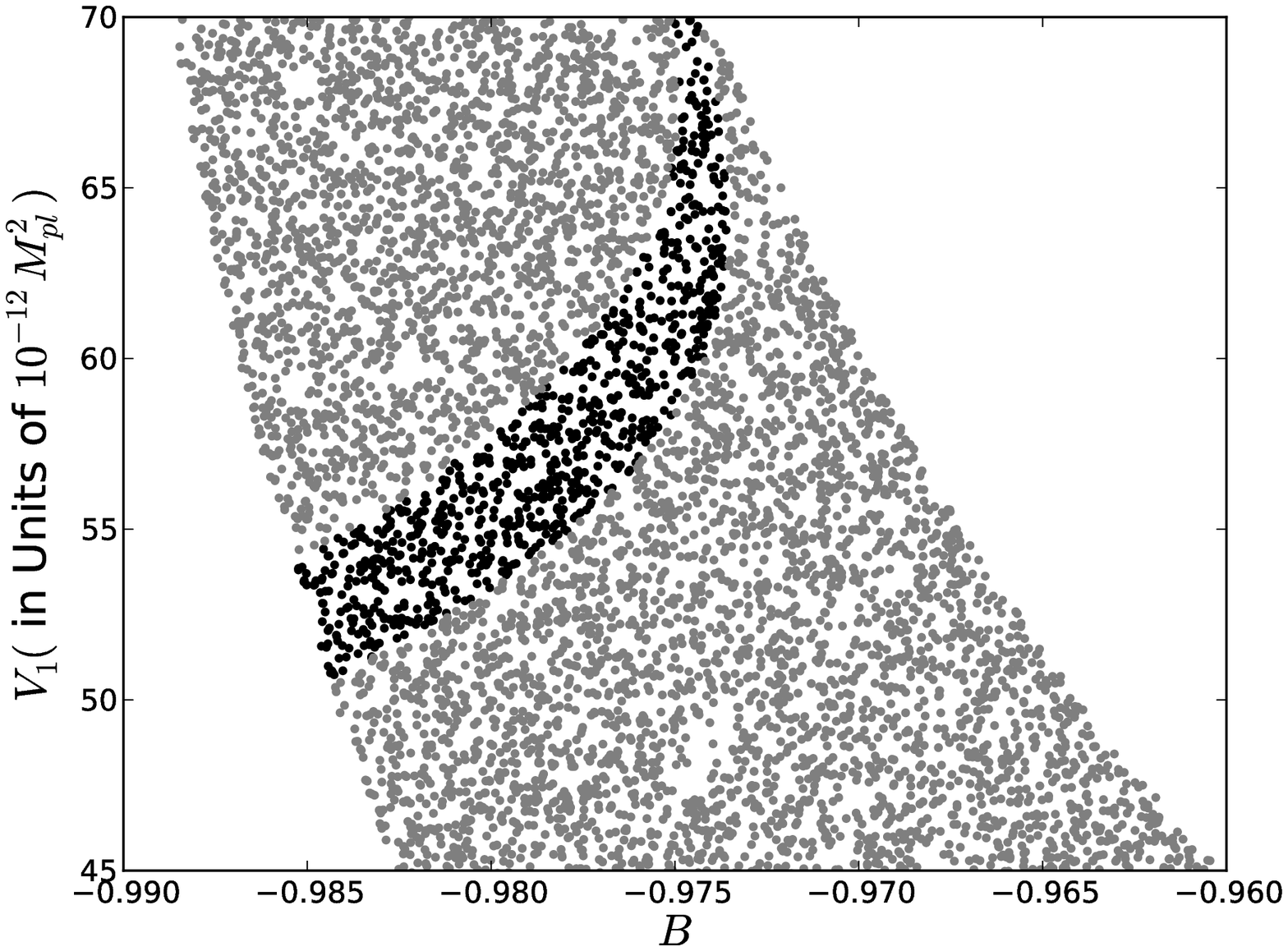,width=7 cm}\\
\epsfig{file=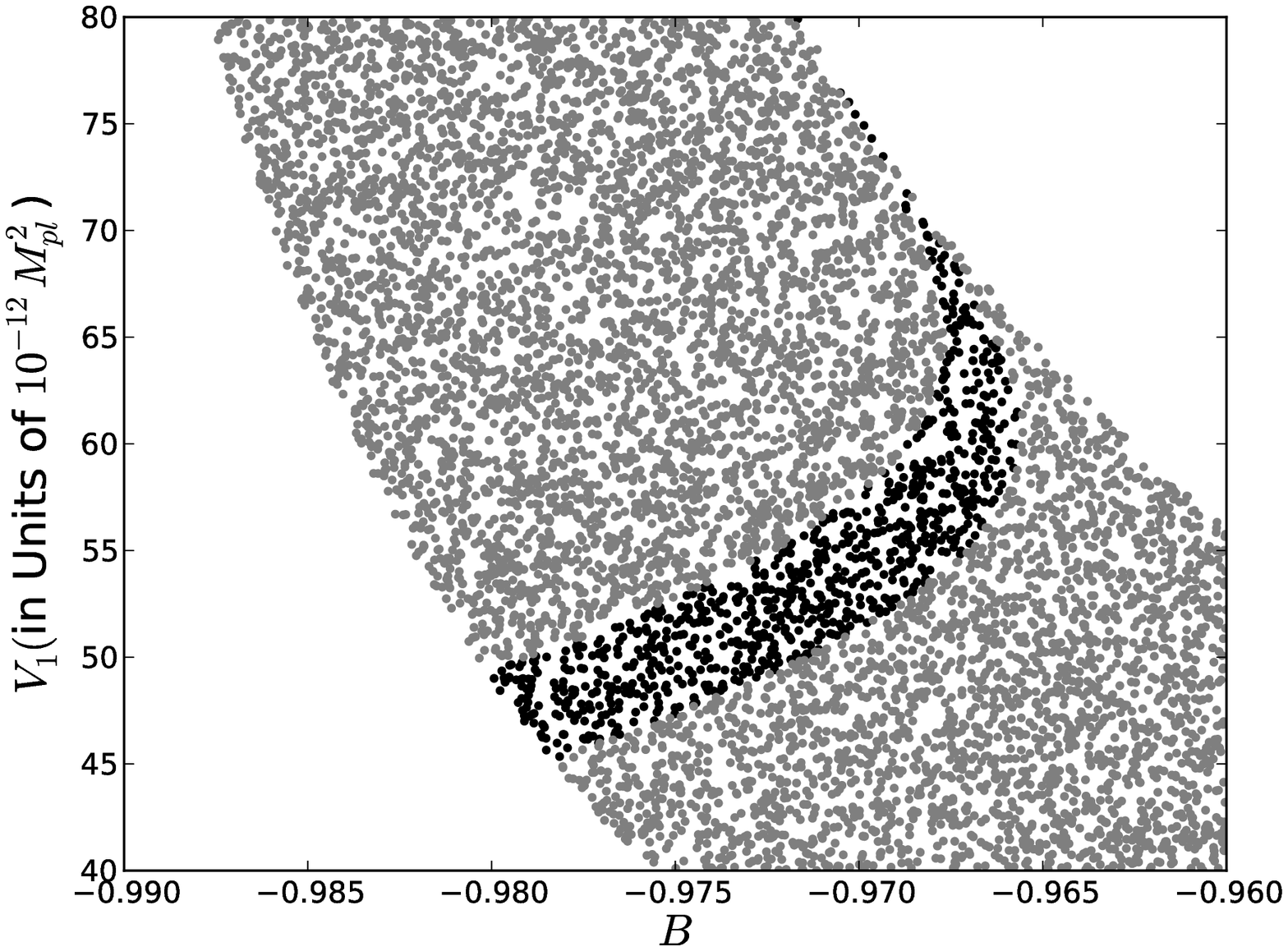,width=7 cm}&
\epsfig{file=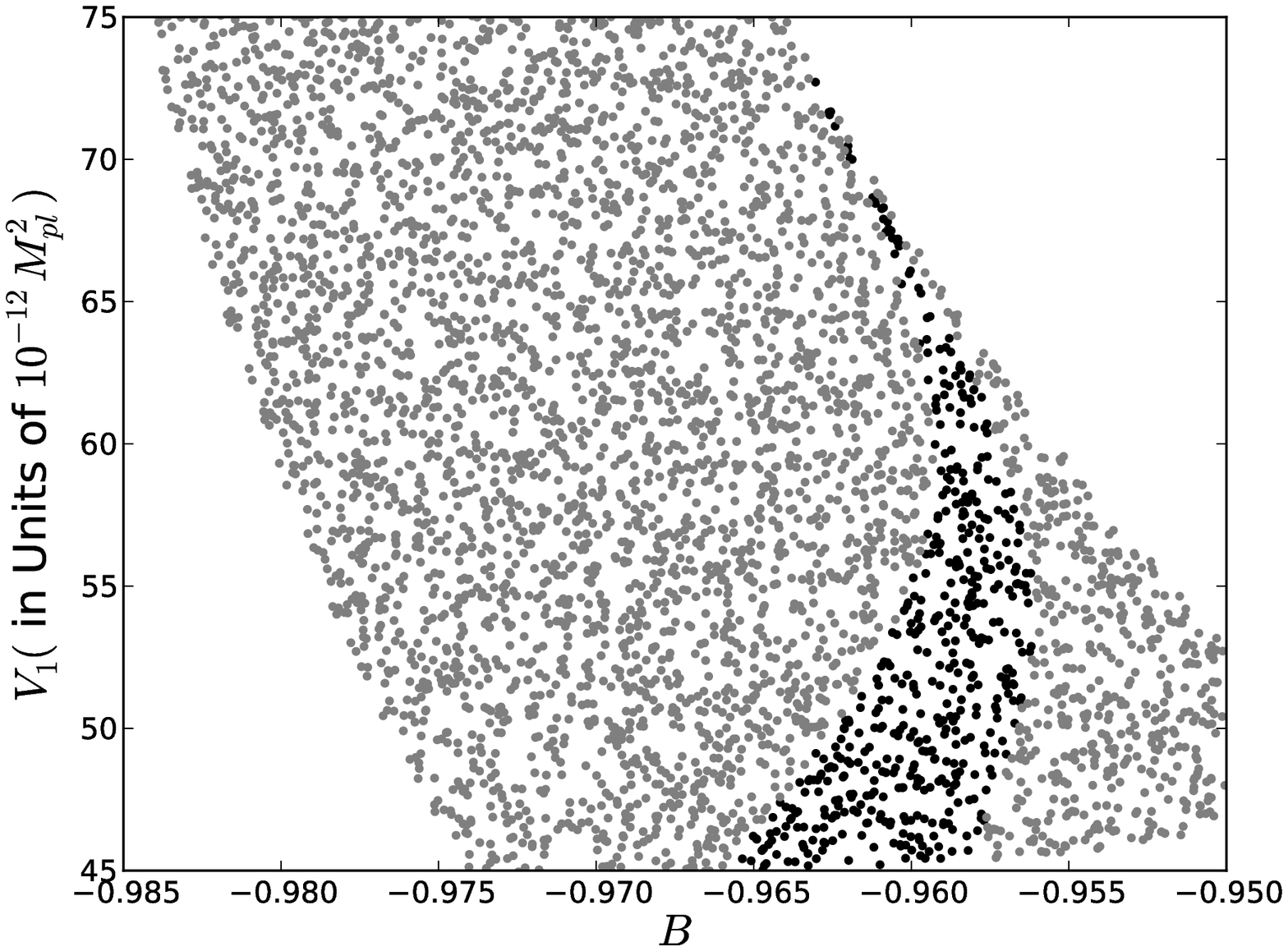,width=7 cm}\\
\end{tabular}
\caption{The allowed region in $B-V_{1}$ plane for $K = 1$. Grey dots represents the points for which inflation successfully happens but do not satisfy planck constraints.  Black dots do satisfy Planck constraints given in eqn (32). Values of $A$ are $A = -0.6 \times 10^{-12} M_{pl}^2$ (top left), $-0.8 \times 10^{-12} M_{pl}^2$(top right), $-1.0 \times 10^{-12} M_{pl}^2$ (bottom left) and $-1.2 \times 10^{-12}$ (bottom right).}
\end{figure*}
\end{center}

To start with, we fix $K = 1$ without any loss of generality ( this is an arbitrary integration constant). To find out the values of 
parameters ( there are three independent parameters e.g $A$, $B$ and $V_{1}$) for which inflation happens and satisfies Planck constraints (eqn (32)), we 
proceed as follows :  for different values of $A$, we choose a range of values for $V_{1}$ and $B$. We choose random points in the given range and see if inflation happens 
with enough number of e-folds and also it ends before the minimum of the potential. 
If the point does not satisfy these two conditions then we discard them otherwise we calculate values of $A_{s}$, $n_{s}$ and $r$ for those points in parameter space and 
check whether they satisfy the constraints given by Planck as mentioned in eqn (32).  The corresponding results are shown in figure (3).

Further in figure 4, we show the allowed regions for our model in the $n_{s}-r$ plane together with the Planck constraints. We show this for two particular values of $A$, e.g $A= -0.6 \times 10^{-12} M_{pl}^2$ and $A=-1.0 \times 10^{-12} M_{pl}^2$. As usual, we fix $K=1$ without any loss of generality. One can see the allowed regions for our model is very much inside the Planck's $68\%$ confidence region.

In the Table I, we give different values of model parameters which satisfies the Planck constraints and the corresponding values of $d_0/l$ and two cosmological constants at the visible and hidden branes.

\begin{center}
\begin{figure*}[t]
\begin{tabular}{c@{\qquad}c}
\epsfig{file=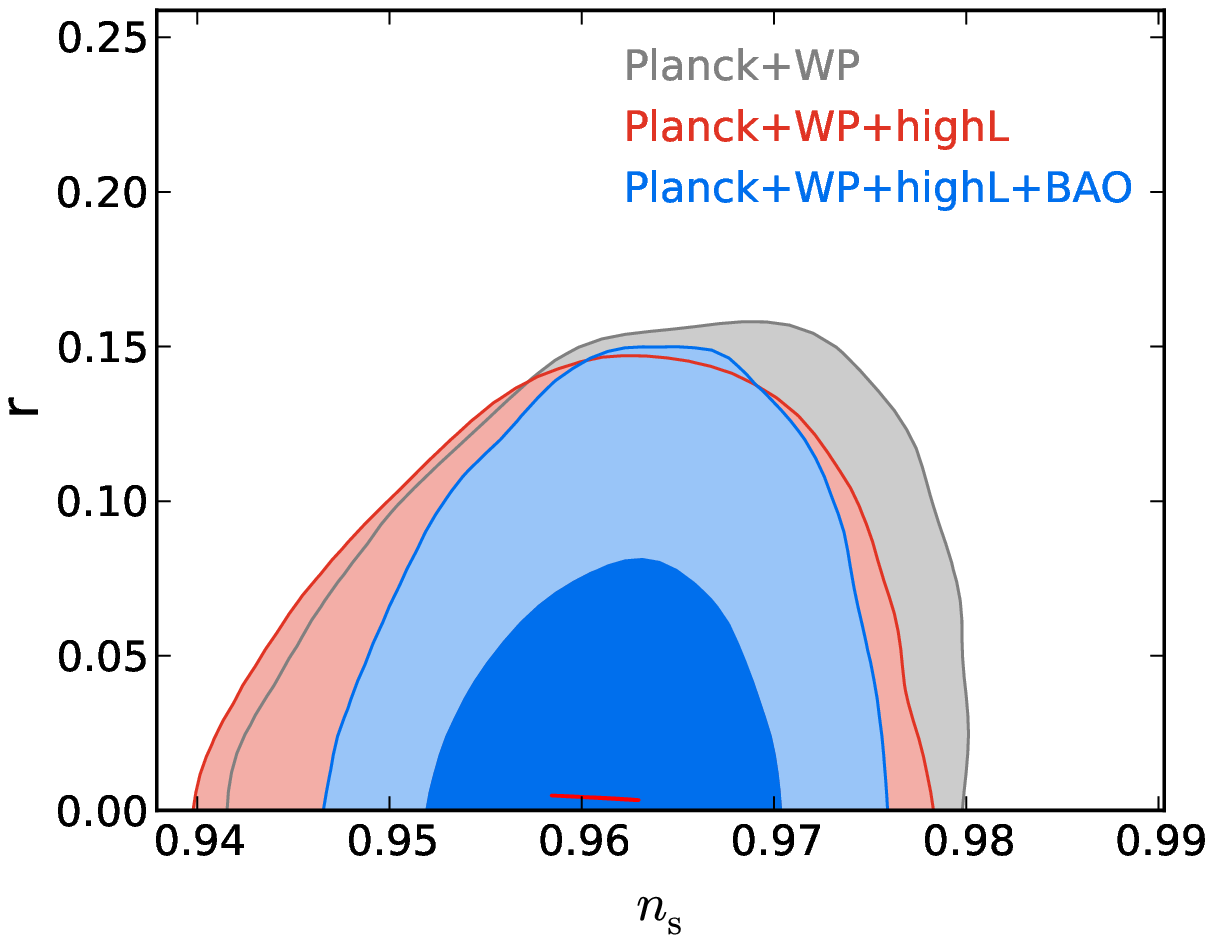,width=7 cm}&
\epsfig{file=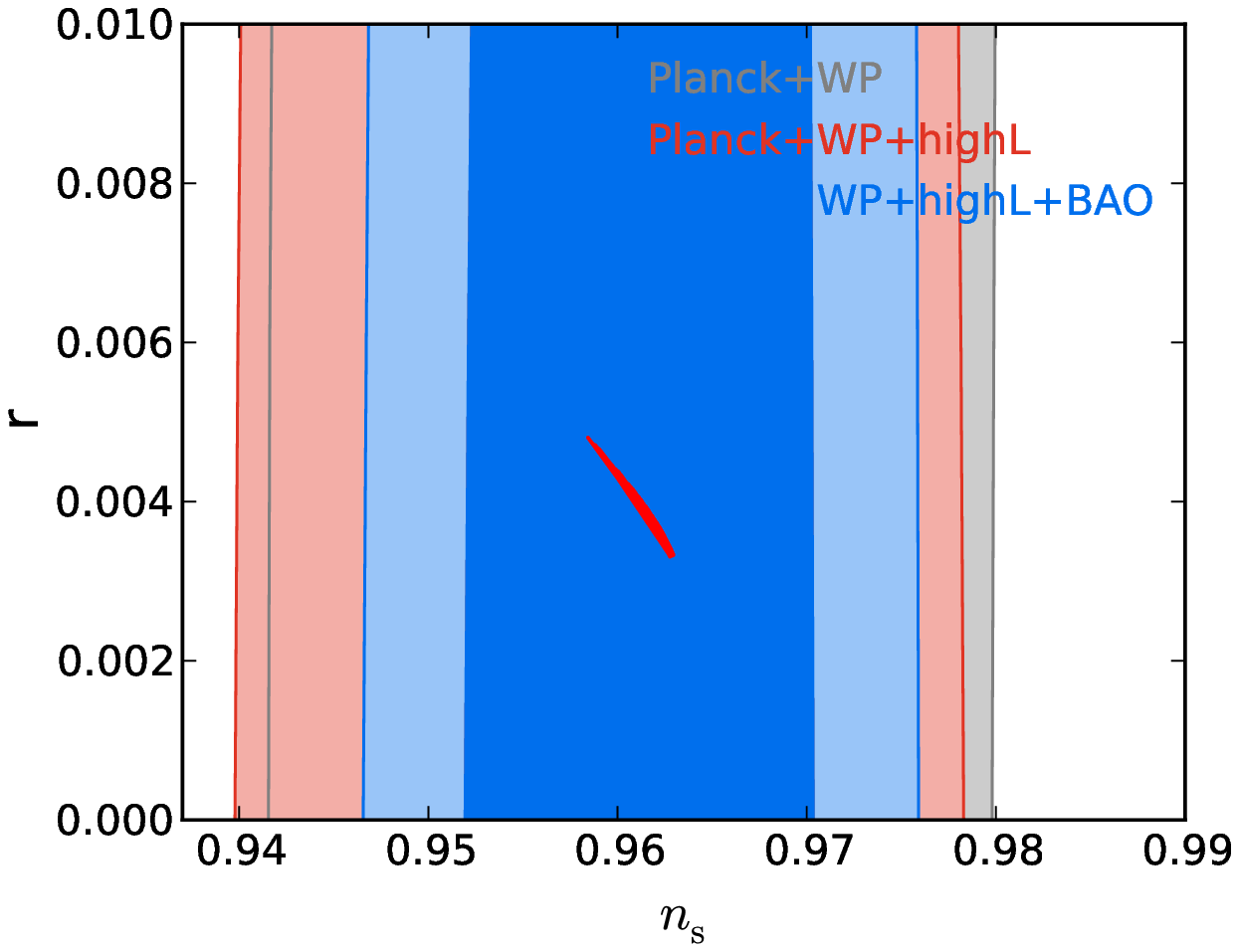,width=7 cm}\\
\epsfig{file=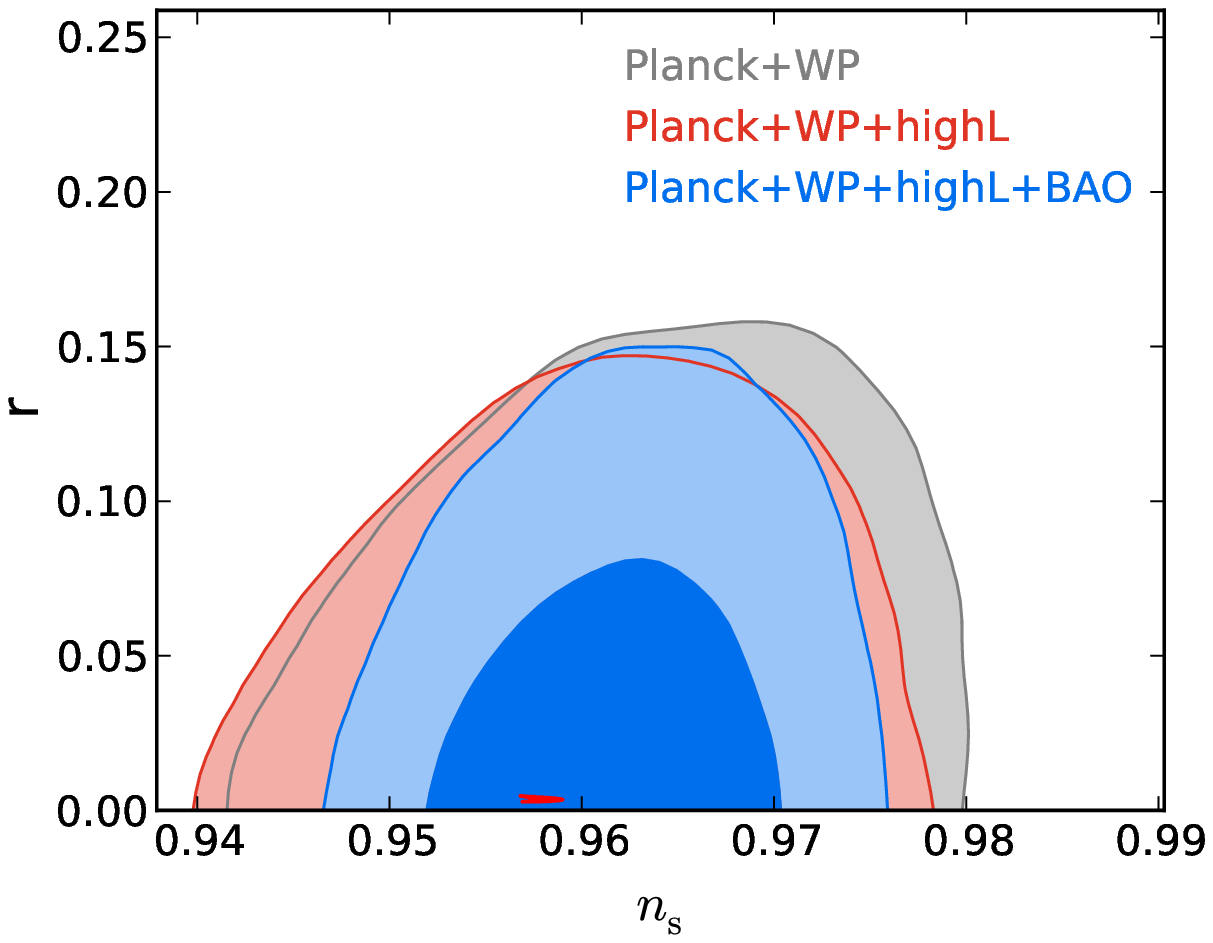,width=7 cm}&
\epsfig{file=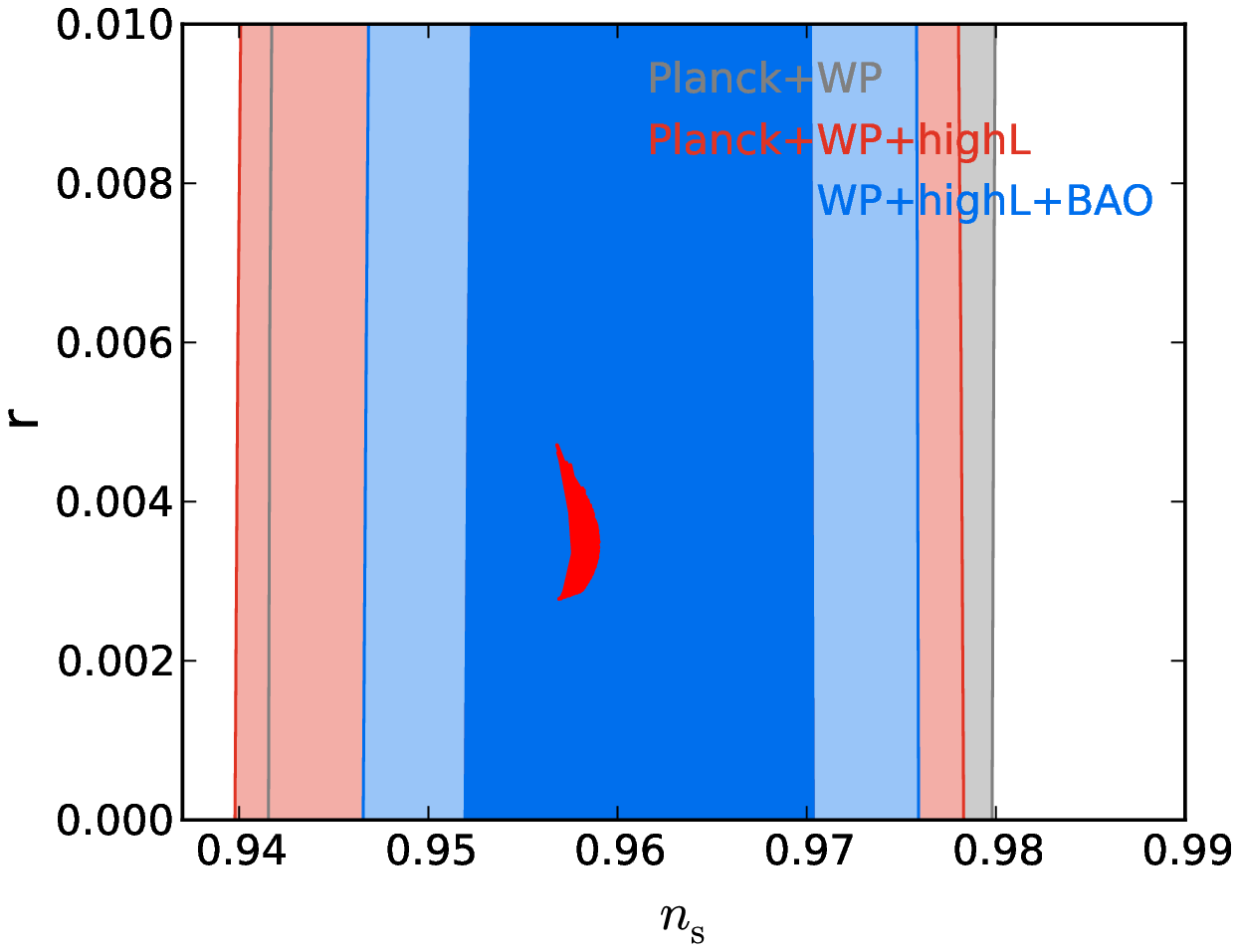,width=7 cm}\\
\end{tabular}
\caption{Top-left Figure: Plot for $n_{s}$ vs $r$ for $A = -0.6 \times 10^{-12} M_{pl}^2$. $K = 1.0$. Bottom-left Figure: Same as above but with $A = -1.0 \times 10^{-12} M_{pl}^2$. In both figures, $N=55$. The right figures are enlarged version of the above figures. The red regions are same as the black regions in Figure 3 for corresponding $A$ values.}
\end{figure*}
\end{center}

\begin{table}[h]
\begin{tabular}{|c|c|c|c|c|c|c|c|c|}
\hline
\multicolumn{3}{|c|}{Model Parameters}                      & \multicolumn{3}{c|}{Cosmological Parameters} & \begin{tabular}[c]{@{}c@{}}Corresponding\\ Brane Distance\end{tabular} & \multicolumn{2}{|c|}{Cosmological constants}  \\ \hline
\begin{tabular}[c]{@{}c@{}}$A$\\$($in units of\\ $10^{-12} M_{pl}^2)$\end{tabular} & B      & \begin{tabular}[c]{@{}c@{}}$V_1$\\$($in units of\\ $10^{-12} M_{pl}^2)$\end{tabular} & $n_s$     & r         & $\ln(10^{10}A_s)$    & $d_0/l$	& \begin{tabular}[c]{@{}c@{}}$\Lambda_1$\\$($in units of\\ $10^{-12} M_{pl}^4)$\end{tabular}	& \begin{tabular}[c]{@{}c@{}}$\Lambda_2$\\ $($in units of\\ $10^{-12} M_{pl}^4)$\end{tabular}                                                                 \\ \hline
-1.2                  & -0.96  & 50                         & 0.955     & 0.003     & 3.09                 & 0.02041      & 57.6 & -60                                                               \\ \hline
-1.0                  & -0.97  & 55                         & 0.959     & 0.003     & 3.09                 & 0.01523      & 64.7 & -66.7                                                      \\ \hline
-0.8                  & -0.98  & 55                         & 0.961     & 0.003     & 3.08                 & 0.01010      & 78.4 & -80                                                       \\ \hline
-0.6                  & -0.985 & 57                         & 0.962     & 0.003     & 3.07                 & 0.00756      & 78.8 & -79.9                                                      \\ \hline
\end{tabular}
\caption{Table for different values of Model parameters $A, B$ and $V_1$  and corresponding values of cosmological parameters $n_s, r$ and $\ln(10^{10}A_s)$. The allowed values for the proper distance between two branes $\frac{d_0}{l}$ and the values of the cosmological constant in the two branes are also listed.}
\end{table}

\section{BICEP2 results for gravity waves}
CMB polarization is one of the most important observational signatures that can give important clues about the physics of very early Universe. 
The E-Mode polarization was first detected by DASI in 2001. But the B-mode polarization which is a clear evidence of primordial gravitational waves generated during inflation has not 
been detected until recently. Just few months before, BICEP2 experiment \cite{bicep2} has announced the detection of the B-mode signal for the CMB polarization thereby confirming the 
existence of the primordial gravitational waves. Their measured value for the tensor-to-scalar ratio $r$ turns out to be  $r=0.2 ^{+0.07}_{-0.05}$ where as  they rule out zero tensor fluctuation 
at $7\sigma$ confidence level. This result brings in huge conflict  with the Planck results on measurement of $r$ which is $r<0.11$. 
But the contribution from Galactic foregrounds to this B-mode signal has been an issue which has to be settled. People have shown that although the BICEP2 data is consistent with $r=0.2$ 
with negligible galactic foreground, it is also consistent with negligible $r$ with significant polarization due to dust \cite{spergel, seljak}. Just recently by using the Planck HFI polarization data for $100$ to $353$ GHz, Adam et al. \cite{planckdust} have shown that polarization signal due to dust over the multipole $40 < l < 120$ is roughly the same as that obtained by BICEP2 over this $ l $ range. This shows that it is entirely possible that the polarization signal BICEP2 has measured is not due to the primordial gravitational wave but due to dust. As this issue is still not settled, we have not considered the BICEP2 results in our analysis. We believe, we should wait till we get the results from ongoing joint analysis of Planck and BICEP2.

However  we should stress that it is difficult to get high value of $r$ in the present model. But added contributions coming from cosmic defects \cite{defects}, primordial 
magnetic fields \cite{mag} as well as cosmic birefringence caused by the coupling between scalar field and the CMB photons through Chern-Simons term \cite{bif} can cause an enhancement to the total contribution for the tensor components.

\section{Conclusion}

To summarize, we study a scalar tensor theory that can explain the moduli stabilisation in the bulk geometry as well as can produce an inflationary Universe in the visible brane which is 
consistent with the recent measurements by Planck experiment \cite{planck}. The scalar tensor theory can naturally arise as an effective 4-dimensional theory through perturbative corrections 
of the brane curvature in a two-brane RS-like set up as obtained earlier by Shiromizu and Koyama \cite{shirkoy}. The potential for the inflaton field is not an ad hoc one but emerges 
from the construction of the model through the effective energy momentum tensor of the modulus field. The dynamics of radion facilitates the inflation and thus offers a natural  
explanation for the origin of inflation.  
Such an inflaton field ( i.e. the radion ) needs to be stabilised and stabilisation of the radion in turn is related to the scalar field sitting at the minimum of the potential. 
To inflict a de-Sitter character to  this minimum, we have to add an uplifting term to the potential which is similar to the de-Sitter lifting by adding fluxes in the KKLT set up. 
We show that one gets enough e-folding in this model 
to solve the flatness and horizon problems. Moreover the primordial fluctuations produced by the inflaton field is consistent with the Planck's 
measurements for $n_{s}$, $r$ and $A_{s}$. Hence the present set up not only provides a viable inflationary scenario which is consistent with the Planck data but also offers a possible
resolution to the modulus stabilisation mechanism concomitantly.

\section*{Acknowledgments}
S.K. thanks the UGC, Govt. of India for financial support. A.A.S. acknowledges the funding from SERC, Dept. of Science and Technology, Govt. of India through the research project SR/S2/HEP-43/2009. SSG thanks CTP, JMI for the local hospitality during his stay where part of the work was done.

\end{document}